\documentclass[superscriptaddress,reprint,
 amsmath,amssymb,
aps,
]{revtex4-1}
\usepackage[dvipdfmx]{graphicx}
\usepackage{amsmath}
\usepackage{amsfonts}
\usepackage{amssymb}
\usepackage{color}

\allowdisplaybreaks[4]

\def\be{\begin{equation}}
\def\ee{\end{equation}}
\def\bea{\begin{eqnarray}}
\def\eea{\end{eqnarray}}

\begin{document}

\title{Towards near-exact solutions of molecular electronic structure: Full coupled-cluster reduction with a second-order perturbative correction}
\author{Enhua Xu}
\affiliation{Graduate School of Science, Technology, and Innovation, Kobe University, Kobe 657-8501 Japan}
\author{Motoyuki Uejima}
\affiliation{Graduate School of Science, Technology, and Innovation, Kobe University, Kobe 657-8501 Japan}
\author{Seiichiro L. Ten-no}
\email{tenno@garnet.kobe-u.ac.jp}
\affiliation{Graduate School of Science, Technology, and Innovation, Kobe University, Kobe 657-8501 Japan}
\affiliation{Graduate School of System Informatics, Kobe University, Kobe 657-8501 Japan}

\begin{abstract}
We introduce a new augmented adaptation of the recently-developed full coupled-cluster reduction (FCCR) with a second-order perturbative correction, abbreviated as FCCR(2).
FCCR is a selected coupled-cluster expansion aimed at optimally reducing the excitation manifold and commutator expansions for high-rank excitations for obtaining accurate solutions of the electronic Sch\"odinger equation in a size-extensive manner.
The present FCCR(2) enables to estimate the residual correlation of FCCR by the second-order perturbative correction $E^{(2)}$ from the complementary space of the FCCR projection manifold.
The linear relationship between $E^{\rm (2)}$ and the energy of FCCR(2) allows accurate estimates of near-exact energies for a wide variety of molecules with strong electron correlation.
The potential of the method is demonstrated using challenging cases, the ground state electronic energy of the benzene molecule in equilibrium and stretched geometries, and the isomerization energy of the transition metal complex, [Cu(NH$_3$)]$_2$O$_2$$^{2+}$.
\end{abstract} 

\maketitle

Devising a reliable model for describing correlated electrons in molecules from first principles is one of the most important challenges in the quantum theory of chemistry and physics.
Coupled-cluster (CC) theory \cite{vcivzek1966correlation,bartlett2007coupled} has enabled both accuracy and efficiency owing to the exponential ansatz of the wave function to cover an encompassing range of applications.
Especially, CC singles, doubles with perturbative triples (CCSD(T)) \cite{raghavachari1989fifth} is referred to as 'the gold standard' of {\it ab initio} theory, and has been widely used to treat single-reference (SR) molecules dominated by the Hartree--Fock (HF) wave functions with high accuracy.
Nevertheless, the CC ansatz for strongly-correlated systems requires higher-rank cluster operators, and is usually formidable within the standard SR framework.
Multi-reference (MR) extensions of CC have partially mitigated this problem \cite{lyakh2012multireference,koehn2013state}, although the prescription of MRCC is yet to be well-established.
In particular, the treatment of a large active space is a crucial issue for practical applications since the majority of MRCC employs a configuration interaction (CI) of complete active space (CAS) as a reference wave function, that is, the full CI (FCI) within the active space.

To ameliorate these features, we recently introduced a novel selected CC approach, the so-called full CC reduction (FCCR) \cite{fccr}.
The exact solution of the $N$-electronic Schr\"odinger equation $\Psi_N$ can be expressed either in the linear (FCI) or exponential (FCC) form, 
\begin{align}
| \Psi_N \rangle &= (1+\hat C_1+\hat C_2+\hat C_3+ \dots +\hat C_N) |0\rangle \label{eq:fci} \\
&= \exp(\hat T_1+\hat T_2+\hat T_3+ \dots +\hat T_N)|0\rangle, \label{eq:fcc}
\end{align}
where, $\hat C_k$ and $\hat T_k$ denote the $k$-fold excitation operators with respect to a suitable Fermi vacuum $|0\rangle$ of a single Slater-determinant.
FCCR makes the best of the sparsity of the exponential ansatz consisting of only connected operators by employing screenings to reduce the projection manifold and commutator operations for higher excitations \cite{fccr}.
The exponential ansatz guarantees the proper scaling of the energy with the system size, and the resulting size-extensivity is the obvious advantage of FCCR over the recent adaptive and stochastic CI approaches \cite{liu2016ici,holmes2016heat,tubman2016deterministic,schriber2016communication,booth2009fermion,cleland2010communications,petruzielo2012semistochastic,ten2013stochastic}.
The projection manifold of FCCR is iteratively updated using the interacting space connected to the primary clusters.
Screening is efficiently performed by taking account of the contribution of exclusion-principle-violating (EPV) terms \cite{fccr}.
\begin{center}
	\begin{figure}[]
		\includegraphics[width=240pt]{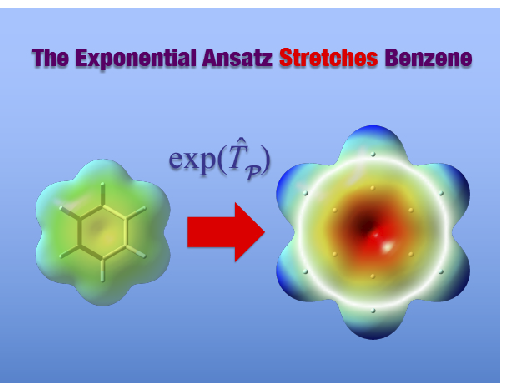}
	\label{fig:scheme}
	\end{figure}
\end{center}
In this Letter, we propose an augmented FCCR with a second-order perturbative correction, termed FCCR(2), inspired by the success of CCSD(T) in the SR case.
We also present the implementation of FCCR(2) along with applications to challenging systems.

The projection manifold of FCCR with high precision contains an enormous number of high-rank cluster operators with small amplitudes, and the iterative treatment of all of them hinders the wide applicability of the method. 
We, therefore, divide the projection manifold and the corresponding cluster operators into those in the primary space ${\mathcal P}$ and secondary space ${\mathcal Q}$.
The FCCR wave function is parametrized within the manifold ${\mathcal P}$,
\begin{align}
\left| \Psi_{\rm FCCR} \right\rangle=e^{\hat T_{\mathcal P}} \left| 0\right\rangle, \label{fccr}
\end{align}
where $\hat {T}_{\mathcal P}=\sum_{\mu\in \mathcal P} t_\mu \hat{a}^{\dagger}_\mu$, $\hat {a}^{\dagger}_\mu$ denotes the excitation operator for the basis $\left| \mu \right\rangle$: $\hat {a}^{\dagger}_\mu \left| 0 \right\rangle=\left| \mu \right\rangle$.
We employ the HF wave function for $\left| 0 \right\rangle$ throughout this work.
The amplitudes $\{t_\mu\}$ and FCCR energy are obtained by solving the working equation \cite{fccr}.
$\hat T_{\mathcal P}$ is comprised of cluster operators with large amplitudes to describe static and the majority of dynamic correlation effects.
Then, the remaining correlation energy from ${\mathcal Q}$ is calculated from a second-order perturbative correction $E^{\rm (2)}$ for the FCCR(2) energy
\begin{align}
E_{\rm FCCR(2)}=E_{\rm FCCR}+E^{\rm (2)}.
\end{align}
The survey of FCCR(2) is depicted in FIG. \ref{fig:scheme}.

The primary space ${\mathcal P}$ is iteratively updated with the aid of the generator space ${\mathcal G}$ as a subset of ${\mathcal P}$.
The present implementation of FCCR employs two screening parameters for the projection manifold, {\it i.e.}, the principal screening threshold $\vartheta_{\mathcal P}$ to select the cluster operators for ${\mathcal P}$ and generator threshold $\vartheta_{\mathcal G}$ for constructing ${\mathcal G}$. The update procedure is given as follows. \\
\hrulefill \\
Update of ${\mathcal P}$: \\
Initialize the primary and generator spaces, ${\mathcal P}_{\rm 0}$ and ${\mathcal G}_{\rm 0}$ to be $\left| 0\right\rangle$, and $n=1$. \\
Step 1 (${\mathcal G}_{\rm n-1}\to{\mathcal P}_{\rm n}$): All single and double (SD) excitations with respect to ${\mathcal G}_{\rm n-1}$ are generated, and those fulfilling $\left|\frac{\sigma_{\kappa}}{\Delta_{\kappa}}\right| \ge \vartheta_{\mathcal P}$ are added to ${\mathcal P}_{\rm n-1}$ to form ${\mathcal P}_{\rm n}$, using the objects $\sigma_{\kappa}$ and $\Delta_{\kappa}$ in Eqs. (\ref{sigma}) and (\ref{Delta}). \\
Step 2 (${\mathcal P}_{\rm n}\to E^{(n)}$): Solve the FCCR working equation in $P\rm_{n}$ for $E^{(n)}$. If $\left|E^{(n)}-E^{(n-2)}\right|<0.5\,\rm{mH}$ and $\left|E^{(n)}-E^{(n-1)}\right|<0.1\,\rm{mH}$, the iteration for ${\mathcal P}$ is terminated. \\
Step 3 (${\mathcal P}_{\rm n}\to {\mathcal G}_{\rm n}$): ${\mathcal G}_{\rm n}$ is updated by adding the determinants in ${\mathcal P}_{\rm n}$ with $|t_\mu|>\vartheta_{\mathcal G}$. Increase $n=n+1$ and return to Step 1.
$\vartheta_{\mathcal G}$ is fixed to be 0.01 in this work. \\
\hrulefill \\
In the above algorithm, ${\mathcal G}$ plays the role of spanning a reference space that can accommodate the excitations necessary to capture strong or non-dynamic correlation in addition to the Fermi vacuum $\left| 0\right\rangle$.
FCCR reduces to CCSD when ${\mathcal G}$ contains only $\left| 0\right\rangle$ and $\vartheta_{\mathcal P}=0$.
However, the calculation with $\vartheta_{\mathcal P}=0$ becomes impractical for large molecules even in the SR case.
Alternatively, we use finite $\vartheta_{\mathcal P}$ in conjunction with the extrapolation of the perturbative correction.
The distribution of the excitation levels of ${\mathcal P}$ or ${\mathcal G}$ is not always continuous, especially for strongly correlated systems, {\it e.g.}, some doubles and quadruples are strongly coupled via two-electron interactions although they are not connected by single excitations.
Therefore, we update the projection manifold using the SD space from ${\mathcal G}$ to construct a compact ${\mathcal P}$ with discrete excitation levels.
The above procedure is quite efficient and reaches the convergence in several iterations.
Operations with negligible contributions arising from the nonlinear structure of the CC working equation are discarded using the operation screening threshold $\vartheta_{\mathcal O}$ in conjunction with the in the EPV form of the equation \cite{fccr}.
A large operation threshold $\vartheta_{\mathcal O}$ may be employed to update $\mathcal P$.

\begin{center}
	\begin{figure}[]
		\includegraphics[width=250pt]{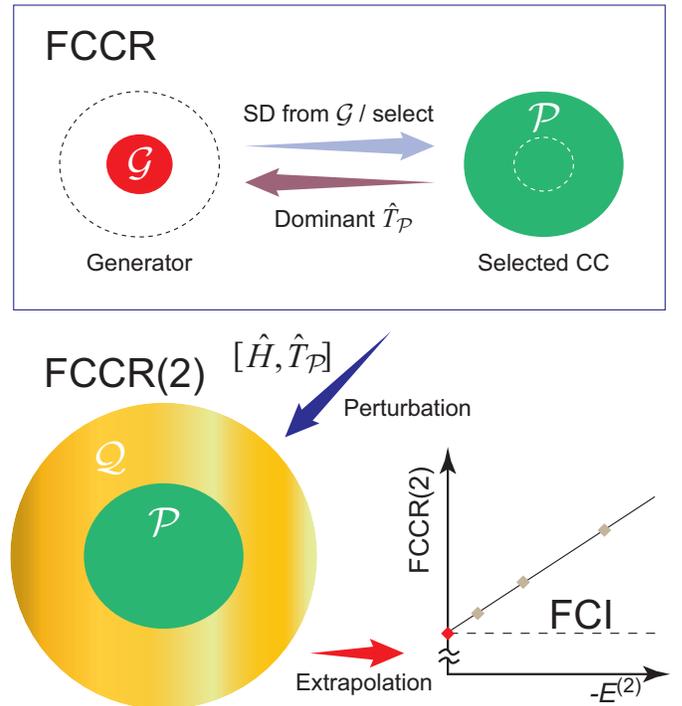}
		\caption{Survey of the FCCR(2) scheme. The primary space ${\mathcal P}$ is iteratively updated with the aid of the the generator space ${\mathcal G}$ in combination with the FCCR calculation. The residual correlation energy from the secondary space ${\mathcal Q}$ is estimated by second-order correction.
		}
	\label{fig:scheme}
	\end{figure}
\end{center}

Once we obtain the amplitudes $\{\hat T_{\mu}\}$ and $E_{\rm FCCR}$, the secondary space ${\mathcal Q}$ is generated for $E^{\rm (2)}$ using the interaction space connected through the single commutator $[ \hat {H},\hat {T}_{\mathcal P} ]$ \cite{fccr}.
The simplest example is all triples generated from $\hat T_{\mathcal P}$ spanning CCSD.
In parallel to the second-order perturbative corrections in SRCC \cite{gwaltney2000second,hirata2001perturbative}, we employ formulas based on the order-by-order expansion in the bi-orthogonal basis for the similarity transformed Hamiltonian, ${\bar H}_{\mathcal P}=e^{-\hat T_{\mathcal P}} \hat {H} e^{\hat T_{\mathcal P}}$.
The second-order correction is expressed as
\begin{align}
E^{\rm (2)}=\sum_{\kappa\in \mathcal Q} \frac{\eta_\kappa \sigma_\kappa}{\Delta_{\kappa}}, \label{eq:e2}
\end{align}
where the objects are
\begin{align}
\eta_\kappa&=\langle 0 | (1+{\hat\Lambda_{\mathcal P}})\bar{H}_{\mathcal P} | \kappa \rangle,
\end{align}
\begin{align}
\sigma_\kappa&=\langle \kappa | \bar{H}_{\mathcal P} | 0 \rangle, \label{sigma}
\end{align}
$\hat\Lambda_{\mathcal P}$ is the left-hand eigen state vector of $\bar {H}_{\mathcal P}$: $\hat\Lambda_{\mathcal P}=\sum_{\mu\in \mathcal P} l_\mu \hat {a}_\mu$ with the amplitudes $\{l_\mu\}$.
We use the simple denominator in the M{\o}ller-Plesset form \cite{hirata2001perturbative} in this particular work,
\begin{align}
\Delta_\kappa= - \langle \kappa | [\hat{F},\hat{a}^{\dagger}_\kappa] | 0 \rangle, \label{Delta}
\end{align}
with the single-electron Fock operator ${\hat F}$ from the HF equation.
In addition, we always monitor the second order correction $E^{\rm (EN2)}$ using the Epstein-Nesbet (EN) type denominator
\begin{align}
\Delta_\kappa^{\rm (EN)}= E_{\rm FCCR}- \langle \kappa | \hat{H} | \kappa \rangle, 
\end{align}
instead of $\Delta_\kappa$ to check the uncertainty from the partitioning.
The computation of the objects, $\left\lbrace \eta_\kappa \right\rbrace$ and $\left\lbrace \sigma_\kappa \right\rbrace$, proportional to the dimension of ${\mathcal Q}$ is considerably more expensive than that of ${\hat T_{\mathcal P}}$ and ${\hat \Lambda_{\mathcal P}}$.
The screenings for the nonlinear operations with the EPV form of the working equations are generalized for these objects by excluding all ${\hat T_{\mathcal P}}$ and ${\hat \Lambda_{\mathcal P}}$ amplitudes relevant to EPV from the screening.
It is found that $E^{\rm (2)}$ is relatively insensitive to the operation threshold $\vartheta_{\mathcal O}$ compared to $E_{\rm FCCR}$, and a larger $\vartheta_{\mathcal O}$ can be employed for $E^{\rm (2)}$. 
The explicit forms of the working equations for $\{ l_\mu \}$, $\{ \sigma_\kappa \}$ and $\{ \eta_\kappa \}$ are detailed in the Supporting Information.
On practicing the second-order correction, the efficiency is increased by excluding insignificant terms of $\mathcal Q$ contributions to $E^{\rm (2)}$.
$\{ \eta_\kappa \}$ containing $\hat a^{\dagger}_{\kappa}$ can be generated much faster than $\{ \sigma_\kappa \}$.
And thus, we calculate $\left\lbrace \eta_\kappa \right\rbrace$ first and then exclude them satisfying $\left|\frac{\eta\rm_{\kappa}}{\Delta\rm_{\kappa}}\right|<\vartheta_{\mathcal F}$ from the summation of Eq. (\ref{eq:e2}).
$\vartheta_{\mathcal F}$ is fixed to be $2\times10^{-7}$ in this work.
These objects are calculated by distributing the loop over $\kappa$ without a large memory requirement, as $\mathcal Q$ is divided into batches.
All FCCR(2) calculations are performed using the \texttt{GELLAN} program \cite{gellan} in a massively parallel implementation.

Recently, extrapolation schemes for selected CI approaches with perturbative corrections have been proposed \cite{tubman2018efficient,li2018fast,tubman2020modern,zhang2020iterative}.
Despite $E^{(2)}$ for FCCR that is radically different from those for CI in the structure and the complexity of the required objects, a linear relationship between $E_{\rm FCCR(2)}$ and $E^{\rm (2)}$ also holds for FCCR(2).
This enables accurate extrapolations of FCCR(2) to near FCI limits with smaller extrapolation distances.

\begin{center}
	\begin{figure}[h]
	\label{fig:n2}
		\includegraphics[width=250pt]{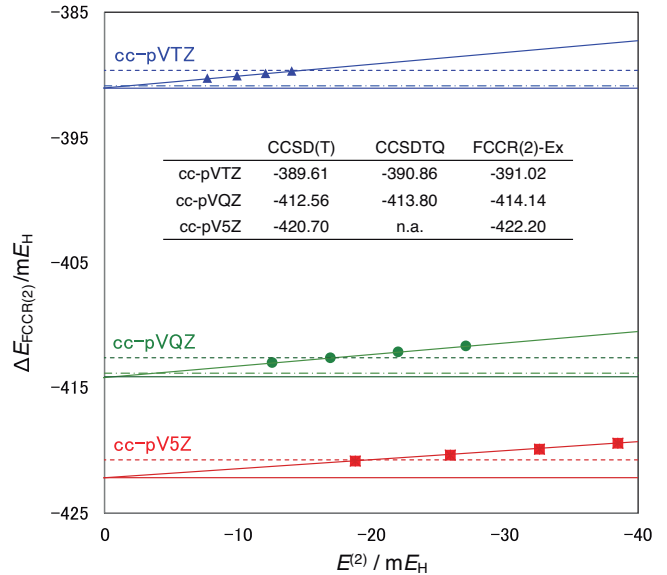}
		\caption{FCCR(2) correlation energies for $\rm N_2$ in the cc-pVXZ basis set (X=T, Q, 5) basis sets at $R_{\rm NN}$=2.068 a.u. using different $\vartheta_{\mathcal P}=5\times10^{-4}$, $4\times10^{-4}$, $3\times10^{-4}$ and $2\times10^{-4}$. The operation thresholds are fixed as $\vartheta_{\mathcal O}=10^{-4}$, $10^{-7}$, and $3\times10^{-6}$ for the update of ${\mathcal P}$, $E_{\rm FCCR}$, and $E^{(2)}$, respectively. The horizontal lines denote CCSD(T) (dashed), CCSDTQ (dash-dotted), and FCCR(2)-Ex (soild) results.}\label{fig:N2}
	\end{figure}
\end{center}

We first present the performance of FCCR(2) for the $\rm N_2$ molecule near the equilibrium bond distance in cc-pVXZ (X=T, Q, 5) basis sets \cite{dunning1989gaussian}.
FIG. \ref{fig:N2} shows the correlation energy of FCCR(2) as a function of $E^{(2)}$ by changing $\vartheta_{\mathcal P}$ in comparison with CCSD(T) and CC singles, doubles, triples, and quadruples (CCSDTQ) \cite{kucharski1991recursive}.
$\Delta E_{\rm FCCR(2)}$ shows nearly perfect linear correlation to $E^{(2)}$ with the root mean square deviations, $\Delta E_{\rm RMSD}=0.007$, 0.017, and 0.027 ${\rm m}E_{\rm H}$ for cc-pVTZ, QZ, and 5Z, respectively.
Consequently, we obtain accurate extrapolations of FCCR(2) by considering the limit of $E^{(2)}=0$, denoted by FCCR(2)-Ex.
The energy differences of CCSD(T), CCSDTQ, and FCCR(2)-Ex are nearly independent of the cardinal number of the basis set, indicating that the majority of the residual correlation beyond CCSD(T) is not short-ranged, such as the electron-electron cusps that can be efficiently described by the F12 ansatz \cite{klopper2006r12,ten2012explicitly,haettig2012explicitly,kong2012explicitly,gruneis2017perspective}.
Note that the FCCR excitation manifold is rather different from the standard CC, {\it e.g.} the generator space ${\mathcal G}$ includes 177 singles and doubles to induce a total 40,919 determinants in ${\mathcal P}$ space  up to quadruples in the cc-pV5Z case.

We apply FCCR(2) to the benzene molecule in the cc-pVDZ basis set \cite{dunning1989gaussian}.
This system was employed in a recent blind test at the equilibrium geometry \cite{eriksen2020ground} for the frozen-core ground state energy of highly-accurate electronic structure methods including CCSDTQ, FCCR(2), many-body expansion FCI (MBE-FCI) \cite{eriksen2018many,eriksen2019many}, selected CI with perturbative corrections \cite{tubman2018efficient,li2018fast,tubman2020modern,zhang2020iterative}, density-matrix renormalization group approach (DMRG) \cite{chan2011density}, adaptive-shift full CI quantum Monte Carlo (AS-FCIQMC)  \cite{ghanem2019unbiasing}, and cluster-analysis-driven FCIQMC (CAD-FCIQMC)  \cite{deustua2018communication}, followed by more recent reports using phase-less auxiliary-field QMC (ph-AFQMC) \cite{lee2020note} and CI of a perturbative selection made iteratively (CIPSI) \cite{loos2020note}.
The total correlation energy in the equilibrium geometry was estimated to be approximately -863 ${\rm m}E_{\rm H}$.
Due to the SR nature of the system, CCSDTQ gives a rather accurate energy, $\Delta E_{\rm CCSDTQ}$=-862.37 m$E_{\rm H}$.
The FCCR(2)-Ex energy amounts to $\Delta E_{\rm FCCR(2)\mathchar`-Ex}=$-862.83 ${\rm m}E_{\rm H}$, which is considered as one of the most accurate results. 
We elaborate on the calculation of FCCR(2).
Nevertheless, a more challenging system is stretched benzene.
Thus, we also examine benzene in a stretched geometry, where all bond distances are extended 1.5 times of the equilibrium geometry.
The screening thresholds are the same as those used for N$_2$.

\begin{table}
	\caption{\label{tab:benzene}
		Dimensions of the ${\mathcal G}$, ${\mathcal P}$, and ${\mathcal Q}$ spaces; $\Delta E^{\rm FCCR(2)}$, and $E^{\rm (2)}$ for the FCCR(2) calculations with $\vartheta_{\mathcal P}=2\times10^{-4}$; the extrapolated energies, $\Delta E_{\rm FCCR(2)\mathchar`-Ex}$ and $\Delta E_{\rm FCCR(EN2)\mathchar`-Ex}$, distances $\Delta E_{\rm dist}$, and RMSD, $\Delta E_{\rm RMSD}$. The energies are in ${\rm m}E_{\rm H}$.}
	\begin{tabular}{lrr}
		\hline
		& \multicolumn{2}{c}{C$_6$H$_6$} \\
		\cline{2-3}
 & equilibrium & stretched \\
		\hline
$N_{\mathcal G}$         &    116  &    1,273 \\
$N_{\mathcal P}$         & 229,842 &  411,628  \\
$N_{\mathcal Q}$         & $6.4\times10^{8}$ &  $5.9\times10^{9}$ \\
$\Delta E_{\rm FCCR(2)}$ & -856.83 & -1,238.63 \\
$E^{\rm (2)}$            &  -35.10 &   -73.55 \\
        \hline
$\Delta E_{\rm FCCR(2)\mathchar`-Ex}$ & -862.82 & -1,252.70 \\
$\Delta E_{\rm FCCR(EN2)\mathchar`-Ex}$ & -863.67 & -1,253.69 \\
$ \Delta E_{\rm dist}$    &   -5.99 &   -14.07 \\
$(\%)^{a}$  &   0.69 &   1.12 \\
$ \Delta E_{\rm RMSD}$                     &   0.03 &    0.02 \\
		\hline
\multicolumn{3}{l}{$^a$Percentage of $\Delta E_{\rm dist}$ in $\Delta E_{\rm FCCR(2)\mathchar`-Ex}$}
			\end{tabular}
\end{table}

Table \ref{tab:benzene} lists the dimensions of the spaces of ${\mathcal G}$, ${\mathcal P}$, and ${\mathcal Q}$ in the spin-orbital basis; $\Delta E^{\rm FCCR(2)}$ and $E^{\rm (2)}$ with the tightest principal threshold; $\vartheta_{\mathcal P}=2\times10^{-4}$ along with $\Delta E_{\rm FCCR(2)\mathchar`-Ex}$, extrapolation distance, $ \Delta E_{\rm dist}$, and $ \Delta E_{\rm RMSD}$ from the extrapolations in the equilibrium and stretched geometries.
In the equilibrium geometry, $\mathcal P$ comprises of excitations from singles to quadruples, and all FCCR amplitudes are less than 0.025.
In contrast, the stretched benzene is a strongly correlated MR system, and the distribution of the excitation manifold of $\mathcal P$ is extended up to septuple excitation levels with 34 amplitudes exceeding 0.1.
Owing to the efficient exponential ansatz, $N_{\mathcal P}$ is only doubled from the equilibrium geometry, although the dimension of ${\mathcal Q}$ for $E^{(2)}$ is increased by one order of magnitude.
The computational cost of FCCR(2) roughly scales as $N_{\mathcal P} N_{\mathcal Q}$.
$N_{\mathcal P}$ are only 0.04 \% and 0.01 \% of the corresponding $N_{\mathcal Q}$ for the equilibrium and stretched benzene, respectively to recover approximately 95\% of the near-exact correlation energy of $\Delta E_{\rm FCCR(2)\mathchar`-Ex}$ using the small excitation manifolds.
FCCR(2) recovers approximately 99\% of the total correlation energy, and the remaining {\it ca.} 1\% is compensated by the extrapolation.
The deviation from $\Delta E_{\rm FCCR(EN2)\mathchar`-Ex}$ (0.85 and 0.99 m$E_{\rm H}$ for equilibrium and stretched) is less than 0.1\% of the total correlation energy.
It appears that the powerful exponential ansatz of $\left| \Psi_{\rm FCCR} \right\rangle$ is effective both for accurate perturbative corrections and extrapolations on top of FCCR in comparison to the selected CI strategies.
We consider the FCCR(2)-Ex energy for the stretched benzene to have similar accuracy as that in the equilibrium geometry.
The types of correlated methods that can approach the FCI solution are open for future researches.

Finally, we present application of our method to a transition metal complex, [Cu(NH$_3$)]$_2$O$_2$$^{2+}$.
It is a well-known bonding motif in biologically relevant molecules related to the peroxo and bis($\mu$-oxo) isomers of a $\rm Cu_{2}O^{2+}_{2}$ core corresponding to the II and III oxidation states of the copper atoms.
The interconversion energy between the two isomers of the [Cu(NH$_3$)]$_2$O$_2$$^{2+}$ model system has been extensively studied using the completely renormalized CC (CR-CC) methods \cite{cramer2006theoretical}, restricted active space second-order perturbation theory (RASPT2) \cite{malmqvist2008restricted}, CAS second-order perturbation theory combined with DMRG (DMRG-CASPT2) with scalar relativistic effect \cite{phung2016cumulant}, and more recent FCIQMC with strongly contracted $N$-electron valence second-order perturbation theory (FCIQMC-sc-NEVPT2) \cite{anderson2020efficient}.
Although most studies predict that the peroxo isomer is more stable than the bis($\mu$-oxo), there remain large inconsistencies in the interconversion energy,
\begin{align}
\epsilon=E_{\rm bis(\mu\mathchar`-oxo)}-E_{\rm peroxo},
\end{align}
{\it i.e.}, $\epsilon_{\rm RASPT2}$=25.2 \cite{malmqvist2008restricted}, $\epsilon_{\rm DMRG\mathchar`-CASPT2}$=22.6 \cite{phung2016cumulant}, and $\epsilon_{\rm FCIQMC\mathchar`-sc\mathchar`-NEVPT2}$=60.0kcal/mol  \cite{anderson2020efficient}.
RASPT2 employed an active space consisting of 24 electrons in 28 orbitals (24,28), and those in DMRG-CASPT2 and FCIQMC-sc-NEVPT2 are (24,24).
We perform FCCR(2) calculations using the same geometries \cite{phung2016cumulant}, and cc-pVTZ basis set for Cu and cc-pVDZ for others keeping the core orbitals frozen.
The screening parameters are identical to those in the previous cases except for $\vartheta_{\mathcal O}$, which is $3\times10^{-7}$ for FCCR and $3\times10^{-5}$ for the perturbative correction to provide a reliable energy difference between the isomers.

\begin{table}[h]
	\caption{\label{tab:cu2o2}
		Dimensions of the spaces and energies of FCCR(2) and statistical measures for FCCR(2)-Ex for the peroxo and bis($\mu$-oxo) isomers of [Cu(NH$_3$)]$_2$O$_2$$^{2+}$.}
	\begin{tabular}{lrr}
        \hline
        &  \multicolumn{2}{c}{[Cu(NH$_3$)]$_2$O$_2$$^{2+}$} \\
		\cline{2-3}
		& peroxo & bis($\mu$-oxo) \\
		\hline
		$N_{\mathcal G}$         &    323  &    407  \\
		$N_{\mathcal P}$         & 625,816 &  643,310 \\
		$N_{\mathcal Q}$         & $3.5\times10^{10}$ &  $4.4\times10^{10}$ \\
		$\Delta E_{\rm FCCR(2)}$ & -1,820.63 & -1,882.74 \\
		$E^{\rm (2)}$            &  -138.23  &  -146.62  \\
		\hline
		$\Delta E_{\rm FCCR(2)\mathchar`-Ex}$  & -1,832.90 & -1,897.44 \\
		$\Delta E_{\rm FCCR(EN2)\mathchar`-Ex}$  & -1,829.94 & -1,895.28 \\
		$\Delta E_{\rm dist}$    &   -12.27 &   -14.70 \\
		$(\%)^{a}$    &   0.67 &   0.77 \\
		$ \Delta E_{\rm RMSD}$                     &   0.16 &    0.10 \\
        \hline
\multicolumn{3}{l}{$^a$Percentage of $\Delta E_{\rm dist}$ in $\Delta E_{\rm FCCR(2)\mathchar`-Ex}$}
			\end{tabular}
\end{table}

Table \ref{tab:cu2o2} shows the details of spaces and energies for the isomers.
The dimensions of ${\mathcal G}$ indicate that both the peroxo and bis($\mu$-oxo) complexes are not as strongly correlated as the stretched benzene, although they require much larger working spaces ${\mathcal Q}$ for perturbative corrections, indicating the presence of high-rank cluster operators in the FCCR calculations.
Indeed, the maximum excitation levels of FCCR are quadruples and quintuples for the peroxo and bis($\mu$-oxo) isomers, respectively.
The final energies of $\Delta E_{\rm FCCR(2)\mathchar`-Ex}$ for the two isomers are considered to be reliable from the relatively small measures for the extrapolations, $\Delta E_{\rm dist}$ and $ \Delta E_{\rm RMSD}$, although the deviation from $\Delta E_{\rm FCCR(EN2)\mathchar`-Ex}$ is somewhat increased compared to the benzene case.

\begin{center}
	\begin{figure}[h]
		\includegraphics[width=250pt]{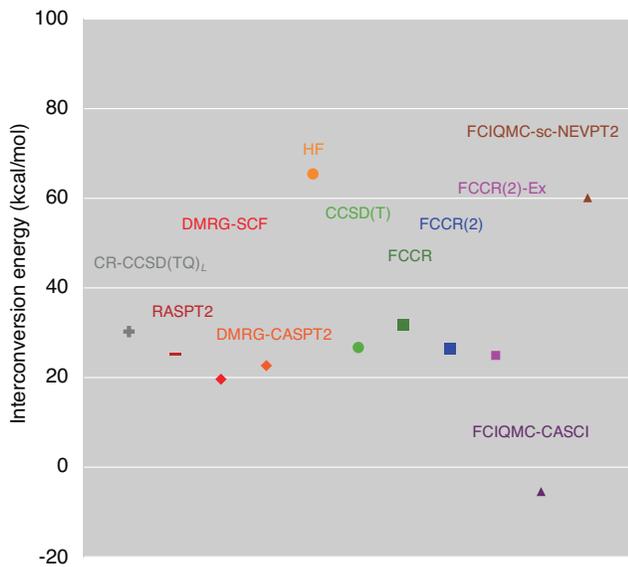}
		\caption{Interconversion energy (kcal/mol) between the peroxo and bis($\mu$-oxo) isomers of [Cu(NH$_3$)]$_2$O$_2$$^{2+}$ of various electronic structure calculations.}\label{fig:cu2o2}
	\end{figure}
\end{center}

FIG. \ref{fig:cu2o2} summarizes the interconversion energies of the isomers from various electronic structure calculations. 
The FCCR interconversion energy is $\epsilon_{\rm FCCR}$= 31.65 kcal/mol.
The second-order perturbative correction reduces this by 5 kcal/mol to yield $\epsilon_{\rm FCCR(2)}$= 26.38 kcal/mol.
Our best estimate of the extrapolated FCCR(2) is $\epsilon_{\rm FCCR(2)\mathchar`-Ex}$= 24.86 kcal/mol in quantitative agreement with RASPT2 employing the relatively large active space (24,28).
Although the FCI limit in the basis set of RASPT2 is not available at this stage, CR-CCSD(TQ)$_L$ also appears to provide accurate result, 30.2 kcal/mol \cite{cramer2006theoretical} in the same basis set.
CCSD(T) gives a reasonable result $\epsilon_{\rm CCSD(T)}$= 26.6 kcal/mol under the same conditions of FCCR, presumably due to error cancelations since the model is too simple for a precise result in general for transition metal complexes.
Overall, the FCIQMC-sc-NEVPT2 estimate employing the HF orbitals is slightly large compared to the other results, indicating that orbital optimization plays an important role in constructing a suitable CASCI reference wave function.
FCCR also uses the HF orbitals without constructing an active space, being capable of treating the orbital relaxation and higher correlation effects via the exponential ansatz to provide the quantitatively accurate interconversion energies.
We cannot pursue further discussions based on the direct comparison due to the difference in the treatment of the basis set and relativistic effects.
Nevertheless, the present benchmark results will be useful for assessing high-quality correlated methods for future studies.

To summarize, we have introduced an augmented version of FCCR with a second-order perturbative correction, FCCR(2) that enables a non-iterative treatment of numerous cluster operators with small amplitudes, which are needed for accurate energetics.
The computation for $E^{(2)}$ can be efficiently parallelized with a small memory requirement with loops over batches dividing the determinants in the secondary space ${\mathcal Q}$.
The advantage of FCCR over a CI expansion is the use of the exponential ansatz that guarantees an accurate result with a small number of parameters for $\hat T_{\mathcal P}$ in a size-extensive manner without constructing an active space.
This feature leads to a powerful FCCR(2) for obtaining near-FCI solutions by reducing the loads on the second-order correction and subsequent extrapolations.
The overall extrapolation distance of FCCR(2)-Ex is only about 1\% of the FCI estimate in this work.
For open-shell systems, the spin symmetry with unrestricted HF references is systematically recovered by increasing the accuracy of FCCR within the present framework.
Further extensions of the FCCR include calculation of excited states and construction of explicitly correlated wave functions.
We leave these research directions for future studies along with discussions on the balance between cost and accuracy against the other methods for strongly correlated systems as partially presented in the case of equilibrium benzene \cite{eriksen2020ground}.

\begin{acknowledgements}
This work was supported by MEXT as a “Priority Issue on Post-K computer (supercomputer Fugaku)” (Development of new fundamental technologies for high-efficiency energy creation, conversion/storage and use) and JSPS Grants-in-Aids for Scientific Research (A) (Grant No. JP18H03900).
We are also indebted to the HPCI System Research project (Project ID: hp150228, hp120278, hp160202, hp170259, hp180216, and hp190175), Academic Center for Computing and Media Studies (ACCMS) of Kyoto University, Education Center on Computational
Science and Engineering (ECCSE) of Kobe university, and the Research Center for Computational Science, Okazaki, Japan for the use of computer resources.
\end{acknowledgements}

\section*{Supporting Information} Details of the working equations and results are collected in Supporting Information.

\bibliography{refs}

\end{document}